\documentclass[conference,9pt]{IEEEtran}
\IEEEoverridecommandlockouts
\usepackage[T1]{fontenc}
\usepackage{cite}
\usepackage{amsmath,amssymb,amsfonts}
\usepackage{graphicx}
\usepackage{textcomp}
\usepackage{xcolor}
\usepackage{booktabs,multirow,tabularx,makecell}
\usepackage[ruled,vlined,linesnumbered]{algorithm2e}
\usepackage{xurl}
\usepackage[hidelinks]{hyperref}
\usepackage[capitalise,noabbrev]{cleveref}
\crefname{figure}{Fig.}{Figs.}
\Crefname{figure}{Fig.}{Figs.}
\crefname{equation}{equation}{equations}
\Crefname{equation}{Equation}{Equations}

\newenvironment{myproblem}[1]
  {\par\noindent\textbf{#1:}\space\ignorespaces}
  {\par}

\DontPrintSemicolon
\SetAlgoLined
\SetKwInOut{Input}{Input}
\SetKwInOut{Output}{Output}
\SetNlSty{textbf}{}{}
\hypersetup{
  pdftitle={Circuit-Guided Exploration of Design Rules and Manufacturing Options for Advanced Technology Nodes},
  pdfauthor={Xinyi Yang, Yuhao Ren, Shuofeng Wang, Chen Shen, Handong Zhu,
    Shouzhen He, Zhouyang Wu, Zhiang Wang, Fan Yang, Keren Zhu},
  pdfsubject={arXiv preprint},
  pdfkeywords={Design-technology co-optimization, standard cell, manufacturing}
}
\ifdefined\pdfminorversion
\else
  
\fi
\begin{document}
\title{Circuit-Guided Exploration of Design Rules and Manufacturing Options for Advanced Technology Nodes}
\author{%
  \IEEEauthorblockN{%
    Xinyi Yang\IEEEauthorrefmark{1}, Yuhao Ren\IEEEauthorrefmark{1},
    Shuofeng Wang\IEEEauthorrefmark{2}, Chen Shen\IEEEauthorrefmark{2},
    Handong Zhu\IEEEauthorrefmark{1},\\
    Shouzhen He\IEEEauthorrefmark{2}, Zhouyang Wu\IEEEauthorrefmark{1},
    Zhiang Wang\IEEEauthorrefmark{1}, Fan Yang\IEEEauthorrefmark{1}, and
    Keren Zhu\IEEEauthorrefmark{1}\thanks{Keren Zhu is the corresponding author
      (\href{mailto:krzhu@fudan.edu.cn}{\nolinkurl{krzhu@fudan.edu.cn}}).}}
  \IEEEauthorblockA{%
    \IEEEauthorrefmark{1}Fudan University, Shanghai, China\qquad
    \IEEEauthorrefmark{2}PFTN Semiconductors Co. Ltd., Shanghai, China}
}

\maketitle
\thispagestyle{empty}
\suppressfloats[t]

\begin{abstract}
Design rules and fabrication choices jointly determine the manufacturability
and circuit performance of advanced-node technologies. Evaluating only
predefined technology configurations leaves many of these trade-offs
unexplored. We present a design-technology co-optimization (DTCO) framework
that uses feedback from circuit implementation to explore design rules and
manufacturing options jointly. For each candidate, process simulation models
how the chosen fabrication strategy realizes the design rules. The framework
combines yield and manufacturing cost estimates with power, performance, and
area estimates from standard-cell characterization and circuit implementation.
Multi-fidelity, multi-objective Bayesian optimization uses these evaluations
to select subsequent candidates. On the EPFL benchmarks, the selected
configurations outperform both technology baselines for 73.9\% of 1,001
objective-weight settings. At comparable evaluation costs, the search achieves
94.3\% higher hypervolume than random search and 103.4\% higher hypervolume
than single-fidelity Bayesian optimization.
\end{abstract}


\section{Introduction}
\begin{figure}[t]
    \centering
    \includegraphics[width=\columnwidth]{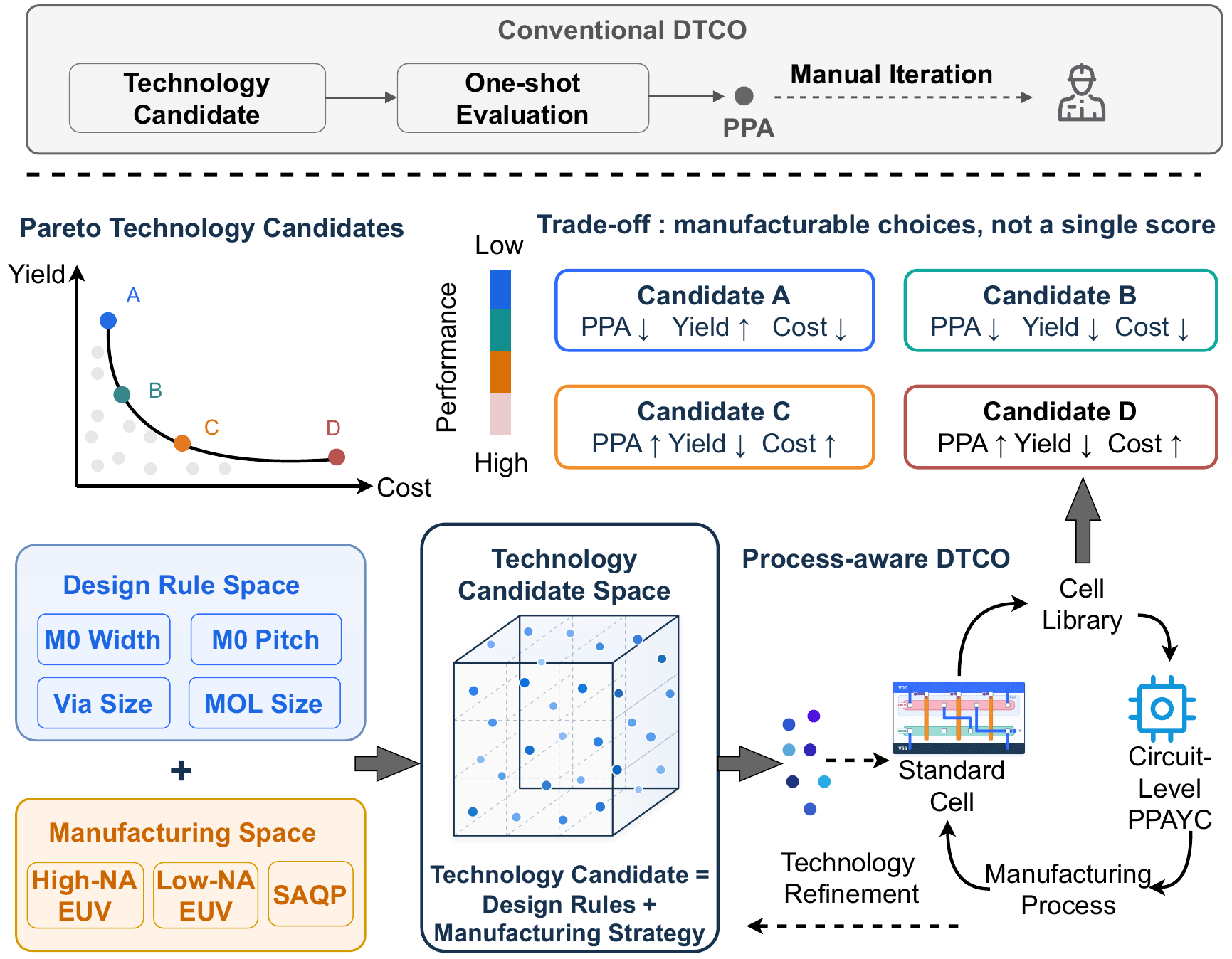}
    \caption{Conventional versus closed-loop DTCO methodologies.}
    \label{fig:pathfinding-comparison}
\end{figure}
As semiconductor technologies enter the sub-5nm era, defining the next technology node has 
become increasingly challenging. As the benefits of conventional device scaling become 
harder to sustain, technology decisions in the middle-of-line (MOL) and 
back-end-of-line (BEOL), including interconnect geometries, routing resources, 
and manufacturing strategies, play an increasingly important role in determining 
PPAYC~\cite{irds2021more-moore,park2021DTCOchallenges}. 
In particular, MOL/BEOL dimensions strongly influence parasitic resistance and capacitance, 
which can substantially limit circuit performance~\cite{fan2017middle,irds2021more-moore}.

The feasibility and cost of these geometric choices also depend on the manufacturing routes 
used to realize them. For example, advanced nodes selectively use extreme 
ultraviolet (EUV) lithography for the most patterning-intensive layers while 
retaining deep ultraviolet (DUV) lithography for less demanding layers~\cite{asmlEuvDuv}. 
Similar geometric requirements may be realized through different patterning routes with 
substantially different mask counts, process complexity, yield maturity, and manufacturing 
cost~\cite{samsung2018euv}. These process-level trade-offs
motivate treating technology pathfinding as the joint exploration of
technology design rules and manufacturing strategies over a broad
candidate space, rather than as the optimization of isolated process
parameters.

DTCO now spans multiple abstraction levels, yet existing methodologies remain fragmented across them. \textbf{At the lower levels}, many studies focus on device optimization, process modeling, or compact-model generation without propagating technology decisions to circuit-level implementation~\cite{park2024TCAD-DTCO,li2024overviewsemidevicemodel,huang2024resnet-model-dtco,peng2025phimo}. Other works emphasize standard-cell layout and design enablement while treating fabrication processes and manufacturing strategies as predefined constraints~\cite{seo2017pin-access,lee2020SMTCell,li2022nctucell}. More comprehensive DTCO frameworks propagate predictive technology assumptions through standard-cell libraries and circuit implementation, but typically evaluate predefined technology configurations rather than jointly exploring their design rules and manufacturing realizations~\cite{choi2023probe3}. \textbf{At higher abstraction levels}, system-technology co-optimization (STCO) attempts technology exploration toward workloads, system architectures, chiplets, heterogeneous integration, and advanced packaging~\cite{pal2024STCO,samavedam2020future,ren2025orthrus}, while generally abstracting the detailed manufacturing realization of individual technology candidates. Consequently, a gap remains between manufacturing-aware technology construction and circuit-level design feedback. A more detailed discussion of related work is provided in \Cref{sec:related-work}.

To bridge this gap, we propose a process-aware, closed-loop DTCO framework
for technology pathfinding, as illustrated in \Cref{fig:pathfinding-comparison}.
The framework automatically explores candidates that combine configurable
technology options with manufacturing strategies. Process simulation
instantiates each candidate as a manufacturable technology implementation,
which is then evaluated through an end-to-end cell- and circuit-level flow.
Multi-fidelity, multi-objective Bayesian optimization (MFMOBO) uses the
resulting PPAYC feedback to select subsequent candidates. This feedback loop
progressively identifies a Pareto frontier of manufacturable technology
solutions.

The major contributions of this work are summarized as follows.
\begin{itemize}

\item We develop an end-to-end closed-loop DTCO workflow integrating
process simulation, library characterization, circuit implementation,
PPAYC evaluation, and iterative technology refinement.

\item We formulate technology pathfinding as joint multi-objective
optimization of technology design rules and manufacturing strategies,
identifying an empirical set of manufacturable candidates nondominated
across the five PPAYC objectives.

\item We develop a process-aware construction flow that converts
technology candidates into manufacturable 3D standard-cell structures
through realistic fabrication modeling, enabling parasitic extraction
and manufacturing-aware yield estimation.

\item Experimental results show that our MFMOBO flow outperforms technology baselines in
73.9\% of 1,001 preference settings and achieves 94.3\%--103.4\% higher
hypervolume than random search and single-fidelity BO at comparable
evaluation cost on EPFL benchmarks.
\end{itemize}

\section{Preliminaries}

This section reviews related DTCO approaches, introduces the patterning
and local-interconnect choices, and formulates the technology pathfinding
problem. Table~\ref{tab:key-notation} summarizes terminology and notation.

\subsection{Related Work}
\label{sec:related-work}
Representative DTCO frameworks span different abstraction levels. At the device and memory-cell levels, Karner et
al.~\cite{karner2021variability} integrate variability modeling based on technology computer-aided design (TCAD), compact-model extraction, full-cell parasitic extraction, and SPICE simulation to evaluate N3 FinFET and nanosheet SRAM technologies. However, the flow terminates at FET and SRAM-cell evaluation, without standard-cell library generation or block-level
circuit implementation. 

At the standard-cell level, iDTCO~\cite{kwon2018iDTCO} incorporates realistic three-dimensional process emulation, parasitic extraction, and manufacturing-aware yield analysis. However, its evaluation is primarily confined to individual standard cells and does not propagate
technology decisions through complete circuit implementation or use circuit-level feedback to refine technology candidates. At the circuit level, PROBE3.0~\cite{choi2023probe3} integrates configurable process-design-kit (PDK) generation, library characterization, and physical design for block-level power, performance, area, and cost evaluation. Nevertheless, it evaluates predefined predictive technology configurations while abstracting the detailed manufacturing processes, process-dependent structures, yield, and fabrication cost associated with each candidate.

Therefore, existing approaches capture only portions of the
process--cell--circuit evaluation chain. They do not jointly construct manufacturable technology candidates and iteratively refine them using circuit-level PPAYC feedback.

\subsection{Patterning}
Patterning technology determines the minimum manufacturable feature pitch and is a key enabler of technology scaling. Representative options include 193-nm immersion (193i) lithography with self-aligned double patterning (SADP) or self-aligned quadruple patterning (SAQP), as well as low- and high-numerical-aperture (NA) EUV lithography. SADP improves pattern density by forming self-aligned sidewall spacers around
lithographically defined mandrels, effectively doubling the achievable pattern density compared with a single lithography exposure. Building upon SADP~\cite{SADP}, SAQP~\cite{chen2013SAQP}
introduces an additional spacer formation and mandrel removal cycle, further doubling the pattern density and enabling significantly smaller feature pitches.
Their representative one-dimensional
pitch limits and relative manufacturing costs~\cite{vashishtha2022asap5} are summarized in
Table~\ref{tab:patterning}. SADP is discussed as the technological basis of SAQP but is not included in the evaluated search space.

\begin{table}[tb]
  \centering
  \caption{Representative patterning options.}
  \label{tab:patterning}
  \normalsize
  \renewcommand{\arraystretch}{1.20}
  \setlength{\tabcolsep}{5pt}
  \begin{tabular*}{\columnwidth}{
      @{\extracolsep{\fill}}
      l
      c
      c
      @{}
    }
    \toprule
    \textbf{Patterning option} &
    \textbf{Pitch Limit (nm)} &
    \textbf{Relative cost} \\
    \midrule
    193i SAQP                  & 20 & Medium    \\
    EUV (0.33 NA, 1D)          & 26 & High      \\
    High-NA EUV (0.55 NA, 1D)  & 16 & Very high \\
    \bottomrule
  \end{tabular*}
\end{table}

These options entail different manufacturability--cost trade-offs and can
affect process characteristics. We therefore include the patterning option
as a variable in the technology search space.

\subsection{MOL and BEOL for Standard Cells}
In advanced-node standard cells, the MOL and first BEOL routing layer
(M0) connect device terminals, support intra-cell routing and pin access,
and strongly affect local parasitics, cell height, routability, and
patterning feasibility~\cite{irds2021more-moore,fan2017middle,
miao2015analytical,chehab2021twolevelMOL}. We therefore jointly explore
their geometries and the associated M0 patterning option, and propagate
the resulting effects to circuit-level PPA. Upper-metal layers are held
fixed because their larger pitches and less aggressive dimensions make
them less sensitive to the cell-level patterning trade-offs considered
here. As the densest metal layer, M0 directly couples manufacturability
with cell-level density and performance and is therefore used as the
representative BEOL technology knob.

\begin{table}[tb]
  \centering
  \caption{Technology variables and search ranges (dimensions in nm).}
  \label{tab:technology-variables}
  \normalsize
  \renewcommand{\arraystretch}{1.08}
  \setlength{\tabcolsep}{4pt}
  \begin{tabularx}{\columnwidth}{@{}c c >{\raggedright\arraybackslash}X@{}}
    \toprule
    \textbf{Variable} & \textbf{Range} & \textbf{Meaning} \\
    \midrule
    \(p_{\mathrm{M0}}\) & \(24\text{--}34\) (step 2) & M0 routing pitch \\
    \(w_{\mathrm{M0}}\) & \(12\text{--}22\) & M0 routing width \\
    \(w_{\mathrm{MD}}\) & \(12\text{--}22\) & Source/drain local-interconnect width \\
    \(w_{\mathrm{MP}}\) & \(12\text{--}22\) & Gate/poly-to-metal local-interconnect width \\
    \(w_{\mathrm{V0}}\) & \(14\text{--}18\) & V0 via width \\
    \(w_{\mathrm{V1}}\) & \(14\text{--}18\) & V1 via width \\
    \(l\)               & \(|\mathcal L|=3\) & M0 lithography/patterning option \\
    \bottomrule
  \end{tabularx}
\end{table}

This work focuses on the local-interconnect and lower-metal parameters that
directly affect standard-cell construction. A technology candidate is denoted
by
\begin{equation}
\theta =
\left(
p_{\mathrm{M0}},
w_{\mathrm{M0}},
w_{\mathrm{MD}},
w_{\mathrm{MP}},
w_{\mathrm{V0}},
w_{\mathrm{V1}},
l
\right) \in \Theta,
\label{eq:technology-point}
\end{equation}
where \(\Theta\) is the technology design space and the variables are
summarized in \Cref{tab:technology-variables}. All geometrical variables take
integer values in nanometers. The M0 pitch is sampled with a step of 2~nm,
whereas all width variables are sampled with a step of 1~nm. The lithography
variable \(l\) is categorical. Their admissible values are further restricted
by connectivity, spacing, and patterning-feasibility requirements.

The search ranges in \Cref{tab:technology-variables} are selected
based on expert knowledge of advanced-node technology scaling,
subject to the lithographic pitch limits in \Cref{tab:patterning}
and geometric feasibility constraints.

\subsection{Problem Formulation}

\begin{myproblem}{Multi-Objective Technology Pathfinding Problem}
Find feasible combinations of design rules and manufacturing strategies
in \(\Theta\) that are nondominated in circuit frequency \(F(\theta)\),
power \(P(\theta)\), area \(A(\theta)\), normalized manufacturing cost
\(C(\theta)\), and process-window yield \(Y(\theta)\).
This mixed-variable, constrained, expensive black-box problem has the
highest-fidelity minimization objectives
\begin{equation}
\mathbf{h}(\theta)=
\left[-F(\theta),\,P(\theta),\,A(\theta),\,C(\theta),\,-Y(\theta)\right].
\label{eq:pathfinding-objectives}
\end{equation}
\end{myproblem}

The Pareto set provides multiple candidates with different PPAYC
trade-offs, enabling designers to select suitable technologies according
to application requirements and preference weights.

\begin{table}[t]
  \centering
  \caption{Terminology and notation.}
  \label{tab:key-notation}
  \fontsize{8}{9}\selectfont
  \renewcommand{\arraystretch}{1.00}
  \setlength{\tabcolsep}{4pt}
  \setlength{\aboverulesep}{1.2pt}
  \setlength{\belowrulesep}{1.2pt}
  \begin{tabularx}{\columnwidth}{@{}l >{\raggedright\arraybackslash}X@{}}
    \toprule
    \textbf{Notation} & \textbf{Meaning} \\
    \midrule
    \multicolumn{2}{@{}l}{\textit{Design space and evaluation metrics}} \\
    \(\theta,\Theta\) & Technology candidate and search space \\
    \(F(\theta),D(\theta)\) & Circuit frequency and delay \\
    \(P(\theta),A(\theta)\) & Power and area \\
    \(C(\theta)\) & Normalized manufacturing-cost proxy \\
    \(Y(\theta)\) & Process-window yield proxy \\
    \(\mathbf h(\theta)\) & Physical minimization objective vector \\
    \(g_Y(\theta)\) & Logarithmic yield loss; scoring uses \(1-Y(\theta)\) \\
    \midrule
    \multicolumn{2}{@{}l}{\textit{Multi-fidelity evaluation and rank transfer}} \\
    \(\mathcal F\) & Fidelity set \(\{F_0,F_1,F_2\}\) \\
    \(F_0,F_1,F_2\) & Inverter, expanded-cell, and circuit evaluation fidelities \\
    \(f,j\) & Fidelity and objective indices \\
    \(\rho_{f,j},m_{f,j}\) & Adjacent-fidelity Spearman correlation and matched sample count \\
    \(a_{f,j},t_{f,j}\) & Target-model and rank-transfer reliability weights \\
    \midrule
    \multicolumn{2}{@{}l}{\textit{Algorithm state and batches}} \\
    \(\mathcal D,\mathcal D^+,\mathcal D^-\) & All, feasible, and infeasible observations \\
    \(\mathcal M\) & Fidelity-specific surrogate models \\
    \(B_f,b_f\) & Selected batch and its size at fidelity \(f\) \\
    \midrule
    \multicolumn{2}{@{}l}{\textit{Candidate selection}} \\
    \(\widehat{\mathbf r}_f(\theta),\mathbf q_f(\theta)\) & Fused and optimistic objective-rank vectors \\
    \(\mathbf s_f(\theta)\) & Heuristic uncertainty vector \\
    \(\beta_f\) & Optimism factor \\
    \(p_f(\theta)\) & Estimated yield feasibility \\
    \(\mathcal P_f\) & Observed feasible rank front \\
    \(\operatorname{HV}\) & Hypervolume \\
    \(\alpha_f(\theta)\) & Feasibility-weighted acquisition score \\
    \midrule
    \multicolumn{2}{@{}l}{\textit{Preference scoring}} \\
    \(z_j(\theta)\) & Min--max normalized objective \\
    \(\mathbf w\) & Nonnegative preference weights, summing to one \\
    \(S(\theta;\mathbf w)\) & Weighted score; lower is better \\
    \bottomrule
  \end{tabularx}
\end{table}

\section{Proposed Method}

\subsection{Overview}
\begin{figure*}[t]
    \centering
    \includegraphics[width=1.0\textwidth]{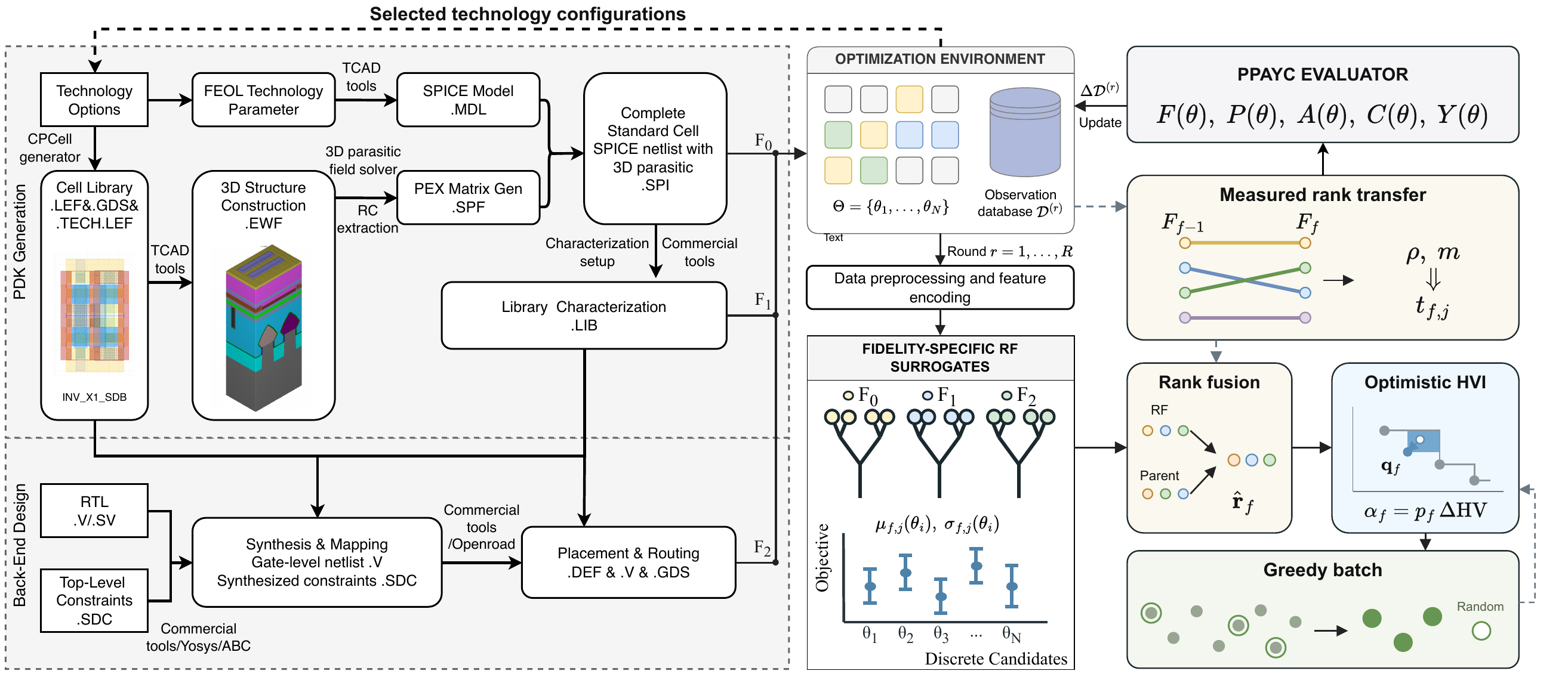}
    \caption{Closed-loop DTCO workflow with nested evaluations,
    fidelity-specific surrogates, measured rank transfer, and greedy batch
    selection using optimistic hypervolume improvement.}
    \label{fig:overview}
\end{figure*}
As shown in \Cref{fig:overview}, the framework takes as input the technology
search space \(\Theta\), specified by \Cref{eq:technology-point} and
\Cref{tab:technology-variables}, together with fixed FEOL parameters,
patterning-specific process recipes,
standard-cell structures, circuit RTL descriptions, and design
constraints. At each iteration, the optimizer selects batches of
technology candidate--fidelity pairs
(\Cref{sec:mf-mobo}).

For each candidate, the technology-construction component instantiates
the corresponding design rules and manufacturing route and generates
material-resolved 3D structures, parasitic and device models, and the
required GDS and LEF views
(\Cref{sec:technology-construction}). Depending on the selected
fidelity, these artifacts undergo cell-level proxy evaluation or library
characterization and complete circuit implementation to obtain
PPA-related observations. In parallel, the manufacturing-cost
(\Cref{sec:cost-model}) and process-variation (\Cref{sec:yield-estimation})
components estimate relative cost and yield, respectively. The optimizer
uses the resulting feasibility and PPAYC observations to guide subsequent
exploration.
Once the evaluation budget is exhausted, the framework reports the
empirical nondominated set from successful \(F_2\) evaluations and
recommends representative points for different preference weights.

\subsection{MFMOBO-Based Technology Pathfinding}
\label{sec:mf-mobo}

To explore technology configurations under a limited evaluation budget, we propose
a multi-fidelity exploration flow that combines cell- and
circuit-level evaluations with iterative PPAYC feedback. The flow uses
lower-fidelity observations to prioritize expensive evaluations and
identify feasible technology trade-offs.
MFMOBO selects candidates and determines which to promote to higher
fidelities.

The evaluations are nested. \(F_0\) evaluates inverter PPA using a
ring-oscillator (RO) test circuit. \(F_1\) expands the cell set by adding
\texttt{NAND2\_X1}, \texttt{NOR2\_X1}, \texttt{AOI21\_X1}, and
\texttt{OAI21\_X1}, with PPA for these additional inverting logic cells
also evaluated using RO test circuits. \(F_2\) completes the standard-cell
library for circuit implementation and circuit-level PPA evaluation.
Manufacturing cost \(C\) and process-window yield \(Y\) are evaluated per
configuration and reused across fidelities.
Candidates with yield below 20\% are not eligible for promotion.
All batches within a round are selected using observations
available at the start of that round. Algorithm~\ref{alg:mf-mobo}
summarizes the model updates, candidate selection, and batch evaluations
performed in each round.

To accommodate the different evaluation scales,
\(\operatorname{FitScoreModel}\) fits independent random forests per fidelity
and objective using fidelity-specific normalized log targets.
Information is transferred through ranking consistency between adjacent fidelities.
Transfer is trusted more when adjacent fidelities agree and have more
paired observations.
\(F_0\) uses predicted ranks; promotions also use measured predecessor PPA ranks,
with transfer reliability
\begin{equation}
 t_{f,j}=\max(0,\rho_{f,j})\min\!\left(1,\frac{m_{f,j}}{10}\right).
\label{eq:rank-transfer}
\end{equation}
Here, \(\rho_{f,j}\) is the Spearman correlation over \(m_{f,j}\) matched
adjacent-fidelity observations; undefined correlations contribute zero.
The fused rank \(\widehat r_{f,j}\) is a weighted average of pool-relative
target-model and measured predecessor ranks, with respective weights
\(a_{f,j}\) (target-model reliability from out-of-fold diagnostics) and \(t_{f,j}\).
Before 12 target observations, predecessor ranks are used
directly; zero combined reliability gives a neutral rank.

Candidate selection balances front expansion, uncertainty, and yield feasibility.
Once sufficiently sampled, \(\operatorname{SelectBatch}\) allocates each
fidelity's batch budget by greedily maximizing optimistic rank-space
hypervolume improvement:
\begin{equation}
\begin{aligned}
 \mathbf q_f(\theta)&=\widehat{\mathbf r}_f(\theta)-\beta_f\mathbf s_f(\theta),\\
 \alpha_f(\theta)&=p_f(\theta)\bigl[
 \operatorname{HV}(\mathcal P_f\cup\{\mathbf q_f(\theta)\})
 -\operatorname{HV}(\mathcal P_f)\bigr].
\end{aligned}
\label{eq:acquisition-function}
\end{equation}
Here, \(\beta_f\) controls optimism, \(p_f\) estimates yield feasibility,
\(\mathbf s_f\) combines heuristic
uncertainty from tree dispersion and fidelity gaps, and \(\mathcal P_f\) is the
observed feasible rank front. Greedy front updates discourage redundant
batch choices.
Auxiliary weighted utility shortlists candidates; approximately \(25\%\)
of each batch provides random exploration.

\begin{algorithm}[t]
\caption{MFMOBO-Based Technology Pathfinding}
\label{alg:mf-mobo}
\SetKwInOut{Input}{Input}
\SetKwInOut{Output}{Output}
\Input{Candidate space $\Theta$, observations $\mathcal{D}$,
       fidelities $\mathcal{F}=\{F_0,F_1,F_2\}$,
       number of rounds $R$, and per-fidelity batch sizes $\{b_f\}$}
\Output{Observed nondominated $F_2$ points}
\For{$r\gets 1$ \KwTo $R$}{
  $(\mathcal{D}^{+},\mathcal{D}^{-})\gets\operatorname{Split}(\mathcal{D})$\;
  $\widetilde{\mathcal{D}}^{+}\gets\operatorname{Prepare}(\mathcal{D}^{+})$\;
  $\mathcal{M}\gets\operatorname{FitScoreModel}(\widetilde{\mathcal{D}}^{+})$\;
  \ForEach{$f\in\mathcal{F}$}{
    $\Theta_f\gets\operatorname{Eligible}(\Theta,\mathcal{D},f)$\;
    \lIf{$f=F_1$}{$\Theta_f\gets\Theta_f\cap\operatorname{Feasible}_{F_0}(\mathcal{D})$}
    \lIf{$f=F_2$}{$\Theta_f\gets\Theta_f\cap\operatorname{Feasible}_{F_1}(\mathcal{D})$}
    $U_f(\theta)\gets\operatorname{Acquire}(\theta,f,\mathcal{M}),
      \quad\forall\theta\in\Theta_f$\;
    $B_f\gets\operatorname{SelectBatch}(\Theta_f,U_f,b_f)$\;
  }
  $\mathcal{D}\gets\mathcal{D}\cup
    \operatorname{Evaluate}\!\left(\bigcup_{f\in\mathcal{F}}B_f\right)$\;
}
\Return{$\operatorname{ParetoFront}(\operatorname{Successful}_{F_2}(\mathcal{D}))$}\;
\end{algorithm}

\subsection{Parameterized GAA Technology Construction}
\label{sec:technology-construction}

We adapt the CPCellUCSD engine
~\cite{abkgroup2026cpcellucsd,cheng2025so3cell} for parameterized GAA
technology construction. Given a technology candidate \(\theta\), the
generator instantiates the corresponding pitches, widths, via
dimensions, routing constraints, and physical-layer mappings to
produce a routed cell layout \(G(\theta)\).

The process emulator combines \(G(\theta)\), fixed FEOL parameters,
and the recipe associated with patterning option \(l\) to execute a
multi-step fabrication flow, including epitaxial growth, implantation,
annealing, deposition, lithography, etching, material removal, and
chemical--mechanical planarization. These operations are repeated as
required by the selected integration route, producing a
material-resolved 3D GAA standard-cell structure, as shown in \Cref{fig:overview}. Invalid
process--geometry combinations are marked infeasible.

The resulting structure is used for 3D field-solver parasitic extraction~\cite{kwon2018iDTCO},
and the extracted RC network is combined with device models to form
the SPICE netlist used for cell characterization. The flow also
produces GDS/LEF views, DRC/LVS reports, and process metadata for
downstream evaluation. The same process flow is evaluated under process variations for the process-window analysis in
\Cref{sec:yield-estimation}.

\subsection{Relative Manufacturing Cost Proxy}
\label{sec:cost-model}
Manufacturing cost is important to technology pathfinding because patterning
routes with similar circuit-level PPA may require substantially different
masks, process steps, and control overhead. We therefore estimate an
industry-informed relative manufacturing-cost proxy for each candidate
\(\theta\) from its M0 patterning route, required process operations, and
pitch-dependent process complexity:
\begin{equation}
C(\theta)=
\bigl(
C_{\mathrm{pat}}+C_{\mathrm{mask}}+C_{\mathrm{exp}}
+C_{\mathrm{dep}}+C_{\mathrm{etch}}+C_{\mathrm{ctrl}}
\bigr)R_{\mathrm{proc}}
\label{eq:cost-model}
\end{equation}
Here, \(C_{\mathrm{pat}}\) is the baseline cost of the selected patterning
route. The terms
\(C_{\mathrm{mask}}\) and \(C_{\mathrm{exp}}\) account for mask preparation
and lithographic exposures, while \(C_{\mathrm{dep}}\) and
\(C_{\mathrm{etch}}\) accumulate the deposition and etch operations required
by spacer formation and pattern transfer. \(C_{\mathrm{ctrl}}\) represents
additional process-control and metrology overhead. Published studies inform
the cost-model structure and relative patterning
costs~\cite{hazelton2008cost,raley2016wafercost}, while operation-level
coefficients are estimated from in-house industrial experience on a common
normalized basis.

\(R_{\mathrm{proc}}\) penalizes increasing process complexity as M0 pitch
approaches route \(l\)'s assumed resolution limit. Pitches below this limit
are infeasible. The penalty is separate from the simulated process-window
yield \(Y(\theta)\), preserving distinct cost and robustness objectives.

\subsection{Process-Window-Based Yield Proxy}
\label{sec:yield-estimation}
\begin{figure}[tb]
    \centering
    \includegraphics[width=\columnwidth]{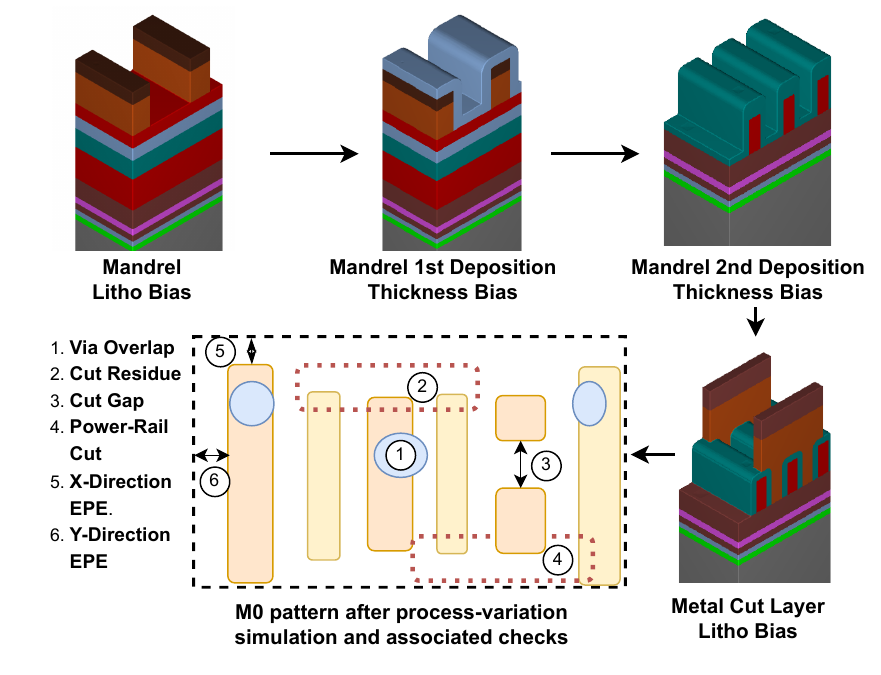}
    \caption{Process-variation simulation flow and the resulting M0 pattern,
    illustrating six checks: via overlap, cut completeness, cut gap,
    unintended power-rail cutting, x-direction EPE, and y-direction EPE.}
    \label{fig:Yield}
\end{figure}

We estimate the process-window yield of each technology point in three steps:
sampling its process variations, emulating the resulting printed M0 geometry,
and testing that geometry against six acceptance checks. The overall flow and
the evaluated failure mechanisms are summarized in \Cref{fig:Yield} and
detailed below.

For each technology point, \(N=100\) Latin-hypercube samples
~\cite{loh1996latin} are drawn from Gaussian distributions of lithography
bias, mask displacement, and deposition thickness. Each sample is processed by
the corresponding lithography emulation flow, and its printed M0 geometry is
compared with the intended color, cut, and V0 geometries. The checks detect
insufficient via coverage, residual metal in cut windows, insufficient cut spacing associated with time-dependent dielectric-breakdown (TDDB) risks~\cite{wong2012TDDB}, unintended power-rail cuts, excessive x- and y-direction edge-placement error (EPE).
They follow established EPE~\cite{ITRS2013Lithography}, cut-process, via-overlap, and reliability
analyses~\cite{irds2023lithoroadmap,zhang2014epe,soussou2025processepe}.

Guided by the cited failure mechanisms, we adopt the following exploratory
acceptance thresholds uniformly across all candidates:
\begin{equation}
\begin{aligned}
|\mathrm{EPE}_{x}| &\leq 3~\mathrm{nm}, &
|\mathrm{EPE}_{y}| &\leq 0.5~\mathrm{nm},\\
A_{\mathrm{residue}} &\leq 2~\mathrm{nm}^{2}, &
d_{\mathrm{cut}} &> 8~\mathrm{nm},\\
r_{\mathrm{via}} &\geq 0.7,
\end{aligned}
\label{eq:yield-acceptance}
\end{equation}
where \(A_{\mathrm{residue}}\), \(d_{\mathrm{cut}}\), and \(r_{\mathrm{via}}\)
denote residual cut-window metal area, minimum neighboring-cut spacing, and
via-overlap ratio, respectively. We also require no unintended power-rail cuts. Missing or invalid simulations are counted as
failures.

We estimate \(Y(\theta)\) as the fraction of the 100 samples passing all
six checks and use
\begin{equation}
g_Y(\theta)=-\log\!\left(\max\!\left(Y(\theta),\epsilon_Y\right)\right)
\label{eq:yield-loss}
\end{equation}
as the yield-loss objective,
where \(\epsilon_Y>0\) handles zero-yield observations.

\section{Experimental Evaluation}
\subsection{Experimental Setup}
We implement MFMOBO in Python~3.12 and execute all evaluations on a Linux
server with two Intel Xeon Silver 4210R CPUs. We use the EPFL benchmark
suite~\cite{epfl_benchmarks_repo} for circuit-level evaluation. Each technology
configuration undergoes library characterization, synthesis, and
place-and-route to obtain post-route delay, core area, and power; cost and
yield use the models in \Cref{sec:cost-model,sec:yield-estimation}.
The search space contains 598,950
configurations, of which 29,066 satisfy the lithography-feasibility and
geometric enclosure constraints.

We use an expert-selected configuration (Expert),
PKP3~\cite{pkp3github,peng2026pkp}, and GT3~\cite{shim2025gt3,gt3github} as
our baselines. For a fair comparison, we evaluate the MOL/BEOL geometries of
PKP3 and GT3 in the same flow, using consistent device models and material
settings.

\begin{table}[t]
\centering
\caption{Preference-dependent scores on EPFL20 and coverage of the
1,001-vector weight grid. Lower scores are better.}
\label{tab:preference-robustness}
\begingroup
\fontsize{8}{9}\selectfont
\setlength{\tabcolsep}{3pt}
\renewcommand{\arraystretch}{1.00}
\setlength{\aboverulesep}{1.2pt}
\setlength{\belowrulesep}{1.2pt}
\begin{tabular*}{\columnwidth}{@{\extracolsep{\fill}}lrrrr@{}}
\toprule
& & & \multicolumn{2}{c}{Best MFMOBO} \\
\cmidrule(l){4-5}
Preference & Expert & GT3~\cite{shim2025gt3} & Point & Score \\
\midrule
Balanced & 0.3653 & 0.6000 & Balanced & \textbf{0.3037} \\
Delay    & 0.4789 & 0.8000 & Fast & \textbf{0.1685} \\
Area     & 0.6435 & \textbf{0.3000} & Compact & 0.4047 \\
Power    & \textbf{0.1827} & 0.8000 & Balanced & 0.2511 \\
Cost     & 0.1827 & 0.8000 & Fast & \textbf{0.1685} \\
Yield    & 0.3389 & 0.3000 & Balanced & \textbf{0.1518} \\
\bottomrule
\end{tabular*}

\vspace{3pt}
\begin{tabular*}{\columnwidth}{@{\extracolsep{\fill}}lrrrr@{}}
MFMOBO versus & Wins & Ties & Losses & Win rate \\
\midrule
Best baseline & 740 & 2 & 259 & 73.9\% \\
\bottomrule
\end{tabular*}

\endgroup
\end{table}

\begin{table*}[t]
  \centering
  \caption{EPFL circuit comparison under the balanced preference.}
  \label{tab:balanced-epfl20}
  \begingroup
  \fontsize{7.5}{8}\selectfont
  \setlength{\tabcolsep}{1.8pt}
  \renewcommand{\arraystretch}{1.00}
  \setlength{\aboverulesep}{1.2pt}
  \setlength{\belowrulesep}{1.2pt}
  \begin{tabular*}{\textwidth}{@{\extracolsep{\fill}}l*{12}{r}@{}}
  \toprule
  Benchmark & \multicolumn{3}{c}{Expert} & \multicolumn{3}{c}{GT3} & \multicolumn{3}{c}{PKP3~\cite{peng2026pkp}} & \multicolumn{3}{c}{MFMOBO-balanced} \\
  \cmidrule(lr){2-4}\cmidrule(lr){5-7}\cmidrule(lr){8-10}\cmidrule(l){11-13}
   & Delay & Area & Power & Delay & Area & Power & Delay & Area & Power & Delay & Area & Power \\
   & (ns) & ($\mu$m$^2$) & (mW) & (ns) & ($\mu$m$^2$) & (mW) & (ns) & ($\mu$m$^2$) & (mW) & (ns) & ($\mu$m$^2$) & (mW) \\
  \midrule
  Adder & 0.836 & 84.59 & 0.150 & 1.036 & 72.49 & 0.187 & 1.195 & 108.61 & 0.209 & 0.946 & 84.59 & 0.186 \\
  Arbiter & 1.281 & 584.65 & 1.396 & 1.487 & 501.40 & 1.534 & 1.466 & 751.49 & 1.679 & 1.285 & 584.65 & 1.510 \\
  Bar & 1.219 & 178.22 & 0.572 & 1.254 & 152.90 & 1.441 & 1.505 & 229.16 & 0.726 & 1.270 & 178.22 & 0.629 \\
  Cavlc & 0.496 & 39.61 & 0.076 & 0.609 & 34.01 & 0.217 & 0.719 & 51.06 & 0.096 & 0.166 & 39.61 & 0.081 \\
  Ctrl & 0.147 & 8.13 & 0.015 & 0.169 & 7.00 & 0.024 & 0.189 & 10.45 & 0.019 & 0.104 & 8.13 & 0.016 \\
  Dec & 0.244 & 21.72 & 0.016 & 0.248 & 18.64 & 0.020 & 0.272 & 27.99 & 0.023 & 0.190 & 21.72 & 0.017 \\
  Div & 69.376 & 3440.53 & 7.011 & 87.113 & 2949.14 & 8.790 & 103.151 & 4423.54 & 9.883 & 86.110 & 3440.53 & 8.521 \\
  Hyp & 907.263 & 15034.66 & 216.951 & 1012.870 & 12888.15 & 253.049 & \multicolumn{3}{c}{N/A\textsuperscript{*}} & 991.052 & 15034.66 & 273.033 \\
  I2C & 0.428 & 72.58 & 0.098 & 0.473 & 62.15 & 0.216 & 0.523 & 93.06 & 0.119 & 0.346 & 72.58 & 0.102 \\
  Int2Float & 0.249 & 12.22 & 0.018 & 0.261 & 10.49 & 0.042 & 0.344 & 15.68 & 0.024 & 0.146 & 12.22 & 0.020 \\
  Log2 & 10.530 & 2060.80 & 16.716 & 11.645 & 1766.51 & 22.910 & 12.281 & 2650.18 & 22.885 & 11.941 & 2060.80 & 19.972 \\
  Max & 4.125 & 211.17 & 1.182 & 3.329 & 180.79 & 3.563 & 6.115 & 271.26 & 1.664 & 5.351 & 211.17 & 1.484 \\
  Mem\_ctrl & 7.731 & 2936.76 & 10.884 & 8.359 & 2516.62 & 17.936 & 7.833 & 3774.13 & 12.238 & 7.935 & 2936.76 & 11.425 \\
  Multiplier & 4.419 & 1577.20 & 10.764 & 4.852 & 1352.32 & 18.388 & 5.961 & 2028.23 & 13.516 & 4.481 & 1577.20 & 11.632 \\
  Priority & 1.984 & 60.09 & 0.134 & 2.510 & 51.56 & 0.311 & 2.960 & 77.45 & 0.197 & 0.218 & 60.09 & 0.165 \\
  Router & 0.526 & 14.24 & 0.017 & 0.587 & 12.19 & 0.020 & 0.683 & 18.17 & 0.023 & 0.551 & 14.24 & 0.020 \\
  Sin & 3.073 & 283.65 & 1.823 & 3.796 & 304.22 & 3.040 & 4.212 & 456.47 & 2.839 & 3.082 & 354.94 & 2.466 \\
  Sqrt & 137.182 & 1600.64 & 25.952 & 157.543 & 1372.26 & 132.833 & 189.359 & 2057.48 & 36.982 & 99.567 & 1600.64 & 31.763 \\
  Square & 3.513 & 1099.71 & 4.941 & 4.087 & 941.88 & 6.703 & 4.471 & 1412.51 & 6.071 & 3.557 & 1099.71 & 5.317 \\
  Voter & 0.874 & 592.22 & 2.610 & 1.120 & 507.87 & 4.228 & 1.269 & 761.40 & 3.844 & 1.063 & 592.22 & 3.266 \\
  \midrule
  Geo. mean (19) & 1.956 & 206.455 & \textbf{0.611} & 2.206 & \textbf{179.145} & 1.103 & 2.562 & 268.484 & 0.815 & \textbf{1.593} & 208.906 & 0.701 \\
  Ratio (19) & 1.000 & 1.000 & 1.000 & 1.127 & 0.868 & 1.804 & 1.310 & 1.300 & 1.333 & 0.814 & 1.012 & 1.147 \\
  \midrule
  Cost proxy & \multicolumn{3}{c}{\textbf{6.4745}} & \multicolumn{3}{c}{7.8440} & \multicolumn{3}{c}{6.8640} & \multicolumn{3}{c}{6.8640} \\
  Yield proxy & \multicolumn{3}{c}{0.750} & \multicolumn{3}{c}{0.800} & \multicolumn{3}{c}{\textbf{0.850}} & \multicolumn{3}{c}{0.800} \\
  Balanced score (19) & \multicolumn{3}{c}{0.2322} & \multicolumn{3}{c}{0.5745} & \multicolumn{3}{c}{0.5396} & \multicolumn{3}{c}{\textbf{0.2090}} \\
  \bottomrule
  \end{tabular*}
  \vspace{3pt}
  \parbox{\textwidth}{\footnotesize
  \textsuperscript{*}PKP3 did not achieve DRC convergence for Hyp within the specified routing-iteration limit; the corresponding PPA values are reported as N/A.}
  \endgroup
  \end{table*}

\subsection{Main Results}
MFMOBO starts with 100 $F_0$ evaluations and accumulates 700, 288, and 30
evaluations at $F_0$, $F_1$, and $F_2$, respectively, over six optimization
rounds. From the feasible $F_2$ evaluations, it identifies an observed
Pareto front.
We select four representative configurations with different objective
trade-offs from this front and keep this set fixed throughout the
preference evaluation. All four configurations, Expert, and GT3 are
evaluated on the same 20 EPFL circuits; PKP3 is excluded from this
full-suite comparison because its Hyp result is unavailable. Delay, area,
and power are aggregated by geometric mean, while cost and yield are
configuration-level proxies.

Unlike the physical objectives \(\mathbf h\) in
\Cref{eq:pathfinding-objectives} and the ranks used for acquisition,
preference scoring uses delay \(D\) and \(1-Y\); \(g_Y\) in
\Cref{eq:yield-loss} is the search's yield-loss transformation.

For preference-based comparison, we min--max normalize
\begin{equation}
\mathbf{x}=(D,A,P,C,1-Y)
\label{eq:preference-objectives}
\end{equation} across the six configurations, keeping the bounds
fixed for all preferences and assigning zero to constant criteria. The
normalized objectives $z_j$ are combined as
\begin{equation}
S(\theta;\mathbf{w})=\sum_{j=1}^{5}w_j z_j(\theta),
\label{eq:preference-score}
\end{equation}
where weights are nonnegative and sum to one; lower scores are better.
Table~\ref{tab:preference-robustness} reports a balanced preference
($w_j=0.2$) and five priority preferences, with weight $0.6$ on one objective
and $0.1$ on each other objective. Balanced (MFMOBO-balanced),
Fast, and Compact denote balanced, speed-oriented, and area-oriented
representatives, respectively.

We further enumerate all 1,001 weight vectors on a 0.1-step grid, including
zero weights. For each vector, the best score among the same four MFMOBO
representatives is compared with the better of Expert and GT3. MFMOBO
outperforms both baselines in $740/1001$ ($73.9\%$) cases, demonstrating
broad preference coverage by the fixed set of representatives on this grid.

\subsection{Ablation Study}
\begin{figure}[t]
  \centering
  \includegraphics[width=0.9\columnwidth]{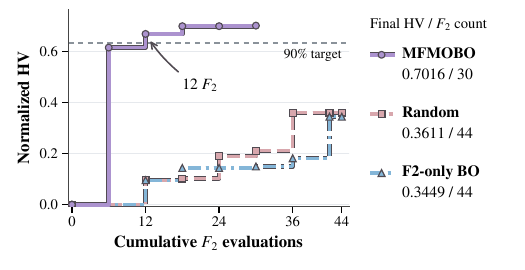}
  \caption{Normalized HV versus cumulative $F_2$ evaluations at batch checkpoints.}
  \label{fig:ablation-hv}
\end{figure}
We compare MFMOBO with random search and $F_2$-only BO to assess the
contribution of multi-fidelity guidance. The latter trains its surrogates
only on $F_2$ observations. Both baselines share 12 initial random $F_2$
configurations and are evaluated through 44 cumulative $F_2$ evaluations.
This budget approximately matches MFMOBO's total evaluation cost:
with a per-evaluation runtime ratio of
\(F_0:F_1:F_2\approx 1:4:130\),
MFMOBO's evaluations across all three fidelities cost approximately as
much as 44 $F_2$ evaluations.
HV is computed from feasible $F_2$ observations of delay, power,
area, cost, and yield loss. All methods and checkpoints share min--max
bounds from the pooled feasible endpoint results, reference point
$(1.1,1.1,1.1,1.1,1.1)$, and normalization by $1.1^5$.

As shown in \Cref{fig:ablation-hv}, MFMOBO reaches HV~0.7016 with 30 $F_2$
evaluations, compared with 0.3611 for random search and 0.3449 for
$F_2$-only BO with 44 evaluations, improving final HV by 94.3\% and 103.4\%,
respectively. It exceeds 90\% of the pooled observed HV at the 12-evaluation
$F_2$ checkpoint; neither baseline reaches this target within 44 evaluations.
These results show higher high-fidelity sample efficiency in this run;
the horizontal axis excludes the additional lower-fidelity evaluation cost.

\subsection{Case Study}
Table~\ref{tab:balanced-epfl20} details the balanced-preference representative
MFMOBO-balanced against Expert, GT3, and PKP3, with equal weights of 0.2 on
delay, area, power, cost, and yield loss. Aggregate PPA and balanced scores
use the common 19 circuits excluding Hyp, with score normalization fixed
over the seven configurations (four representatives and three baselines).
MFMOBO-balanced achieves the lowest balanced score, 0.2090; relative to Expert,
it reduces geometric-mean delay by 18.6\% while increasing area and power
by 1.2\% and 14.7\%, respectively. This illustrates the objective trade-offs
underlying its improved balanced score.

Beyond aggregate metrics, Table~\ref{tab:balanced-epfl20} reveals three broader pathfinding insights for advanced nodes. First, aggressive pitch scaling does not guarantee system-level efficiency: pursuing minimum area alone (as in GT3) incurs substantial power overhead and the highest manufacturing cost, whereas MFMOBO-balanced achieves notable delay and yield gains at nearly identical cell density to Expert through coordinated MOL/M0 sizing. Second, cell-level manufacturability alone cannot ensure block-level routability---PKP3 attains the highest yield proxy yet fails DRC convergence on the routing-intensive \textit{Hyp} circuit, underscoring the necessity of closed-loop $F_2$ physical-design feedback to prevent pin-accessibility bottlenecks. Finally, performance gains vary across circuit topologies, favoring control-dominated logic over interconnect-heavy arithmetic blocks and motivating workload-aware DTCO.

\section{Conclusion}

We presented a process-aware DTCO framework that uses circuit-level PPAYC
feedback to jointly explore design rules and manufacturing strategies.
Experiments on EPFL benchmarks show that MFMOBO outperforms baselines in
73.9\% of 1,001 preference settings and achieves 94.3\% and 103.4\% higher
hypervolume than random search and single-fidelity BO, respectively, under
matched evaluation-cost budgets. These results demonstrate that closed-loop
exploration can move DTCO beyond one-shot evaluation toward systematic and
manufacturing-aware technology pathfinding.

\bibliographystyle{IEEEtran}
\clearpage
\twocolumn
\bibliography{ref/references}
\end{document}